\documentclass[twocolumn]{emulateapj}
\usepackage{apjfonts}

\usepackage{graphicx} 
\usepackage{natbib}

\slugcomment{Accepted for publication in The Astrophysical Journal}

\shorttitle{Progressive star formation in NGC\,3603}
\shortauthors{Beccari et al.}

\begin{document}


\title{Progressive star formation in the young galactic super star cluster NGC\,3603}


\author{Giacomo Beccari\altaffilmark{1}, Loredana Spezzi\altaffilmark{1},
Guido De Marchi\altaffilmark{1}, Francesco Paresce\altaffilmark{2},
Erick Young\altaffilmark{3},
Morten Andersen\altaffilmark{1},
Nino Panagia\altaffilmark{4,5,6},
Bruce Balick\altaffilmark{7}, 
Howard Bond\altaffilmark{4}, 
Daniela Calzetti\altaffilmark{8}, 
C. Marcella Carollo\altaffilmark{9}, 
Michael J. Disney\altaffilmark{10}, 
Michael A. Dopita\altaffilmark{11}, 
Jay A. Frogel\altaffilmark{12}, 
Donald N. B. Hall\altaffilmark{13}, 
Jon A. Holtzman\altaffilmark{14}, 
Randy A. Kimble\altaffilmark{15}, 
Patrick J. McCarthy\altaffilmark{16}, 
Robert W. O'Connell\altaffilmark{17},
Abhijit Saha\altaffilmark{18}, 
Joseph I. Silk\altaffilmark{19}, 
John T. Trauger\altaffilmark{20}, 
Alistair R. Walker\altaffilmark{21}, 
Bradley C. Whitmore\altaffilmark{4}, 
Rogier A. Windhorst\altaffilmark{22} 
}

\altaffiltext{1}{ESA, Space Science Department, Keplerlaan 1, 2200 AG Noordwijk, The Netherlands}
\altaffiltext{2}{INAF - Istituto di Astrofisica Spaziale e Fisica Cosmica, via P. Gobetti, 101, I-40129 Bologna, Italy}
\altaffiltext{3}{NASA--Ames Research Center, Moffett Field, CA 94035}
\altaffiltext{4}{Space Telescope Science Institute, Baltimore, MD 21218, USA}
\altaffiltext{5}{INAF-CT, Osservatorio Astrofisico di Catania, Via S. Sofia 78, 95123 Catania, Italy}
\altaffiltext{6}{Supernova Ltd, OYV \#131, Northsound Road, Virgin Gorda, British Virgin Islands}
\altaffiltext{7}{Dept. of Astronomy, University of Washington, Seattle, WA 98195-1580, USA}
\altaffiltext{8}{Dept. of Astronomy, University of Massachusetts, Amherst, MA 01003, USA}
\altaffiltext{9}{Department of Physics, ETH-Zurich, Zurich, 8093 Switzerland}
\altaffiltext{10}{School of Physics and Astronomy, Cardiff University, Cardiff CF24 3AA, United Kingdom}
\altaffiltext{11}{Research School of Astronomy \& Astrophysics,  The Australian National University, ACT 
2611, Australia}
\altaffiltext{12}{Association of Universities for Research in Astronomy, Washington, DC 20005}
\altaffiltext{13}{Institute for Astronomy, University of Hawaii, Honolulu, HI 96822}
\altaffiltext{14}{Department of Astronomy, New Mexico State University, Las Cruces, NM 88003}
\altaffiltext{15}{NASA--Goddard Space Flight Center, Greenbelt, MD 20771}
\altaffiltext{16}{Observatories of the Carnegie Institution of Washington, Pasadena, CA 91101-1292}
\altaffiltext{17}{Department of Astronomy, University of Virginia, Charlottesville, VA 22904-4325}
\altaffiltext{18}{National Optical Astronomy Observatories, Tucson, AZ 85726-6732}
\altaffiltext{19}{Department of Physics, University of Oxford, Oxford OX1 3PU, United Kingdom}
\altaffiltext{20}{NASA--Jet Propulsion Laboratory, Pasadena, CA 91109}
\altaffiltext{21}{Cerro Tololo Inter-American Observatory, La Serena, Chile}
\altaffiltext{22}{School of Earth and Space Exploration, Arizona State University, Tempe, AZ 85287-1404}

\begin{abstract}

Early release science observations of the cluster NGC3603 with the WFC3 on the refurbished 
HST allow us to study its recent star formation history. Our analysis focuses on stars with 
H$\alpha$ excess emission, a robust indicator of their pre-main sequence (PMS) accreting status. 
The comparison with theoretical PMS isochrones shows that 2/3 of the objects with H$\alpha$ 
excess emission have ages from 1 to 10 Myr, with a median value of 3 Myr, while a surprising 1/3 
of them are older than 10 Myr. The study of the spatial distribution of these PMS stars allows us
to confirm their cluster membership and to statistically separate them from field stars. 
This result establishes unambiguously for the first time that star formation in and around the 
cluster has been ongoing for at least 10-20 Myr, at an apparently increasing rate.

\end{abstract}



\section{Introduction}
\label{intro}

The galactic giant HII region NGC\,3603 located at a distance of
$7\pm1$  kpc~\citep{mof83} is part of the RCW 57 complex SE of Eta Car
in the Carina arm. The  young, bright, compact stellar cluster (HD97950
or NGC 3603YC) lies at  the core of this region and has long been the
center of attention for the relatively numerous population of massive
stars at  its center. The collection of  3 WNL, 6 O3 and numerous late O
type   stars together with a bolometric luminosity of 100 times that of
the  Orion cluster and 0.1 times that of NGC 2070 in the 30 Dor  complex
in the Large Magellanic Cloud (LMC) and a total mass in  excess of
$10^4$\,M$_\odot$  \citep[][hereafter HA08, and references
therein]{har08} places  it squarely in the category of a super star
cluster of the type more  commonly seen in star burst regions of young
galaxies. Thus, it represents a resolved possible prototype of the
building blocks of forming or merging galaxies~\citep[][]{lar09}  and
an excellent opportunity to better understand their complex star
formation history and the physical processes involved~\citep[see, for
example,][]{pet10,moe10,por10}. 

In this context, NGC\,3603YC (from now on referred to as simply NGC 
3603) certainly contains many stars younger  than 3~Myr but it is so far
unclear if it also contains older stars as might be suggested by the
presence in the field of the evolved star Sher\,25~\citep{mel08}, located
$\sim20\arcsec$ north to the cluster center. This exciting
possibility has yet to be established unambiguously because the
latter's  membership in the cluster is still quite uncertain.  The
determination of  the cluster's initial mass function (IMF), of course,
also depends  critically on its formation history as imprinted in the
age distribution that  can only be determined accurately once we
disentangle the various  populations from each other. Results on
this clusterÕs IMF  \citep[HA08;][]{sto06,sun04} may be affected by
imperfect correction for this effect.     

\begin{figure*}  
\centering  
\includegraphics[width=\textwidth]{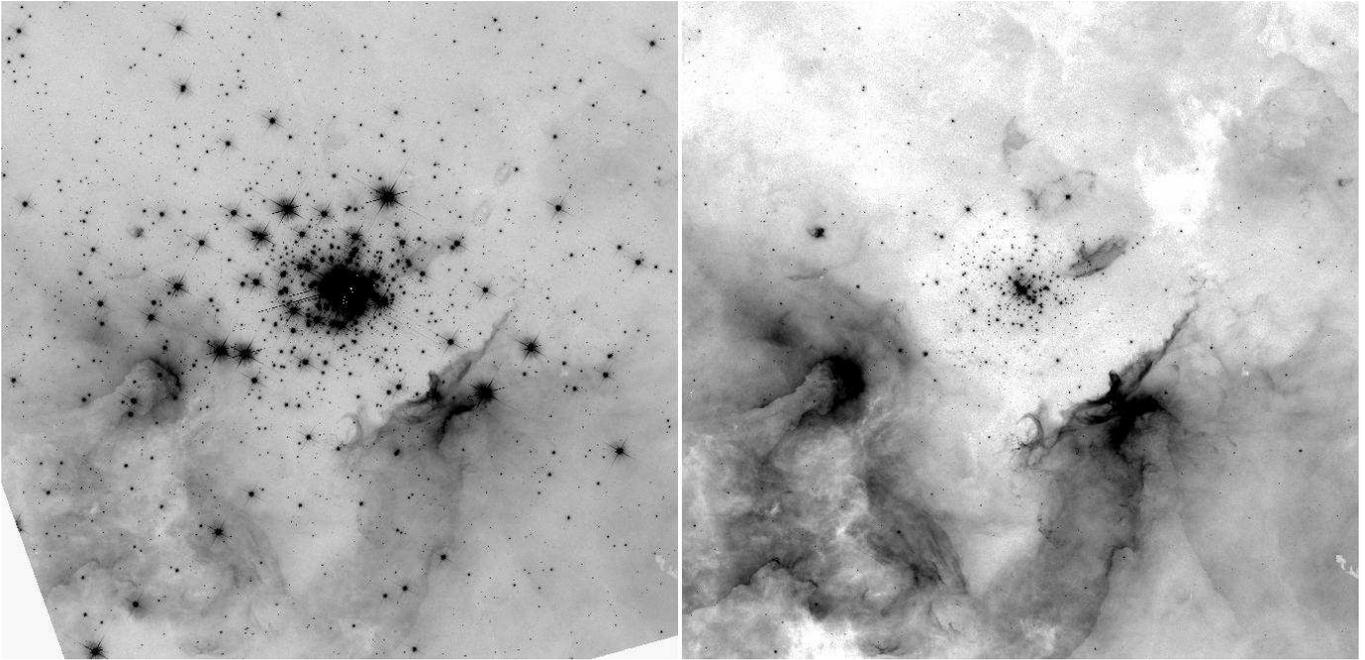}
\caption{WFC3 $2\farcm2\times2\farcm2$ mosaic images in the $F555W$
(left panel) and $F656N$ (H$\alpha$ line; right panel) filters of
NGC\,3603 star forming region as obtained through  PyRAF/MULTIDRIZZLE
package. North is $30\degr$ to the right of the vertical. East is to the
left of North.}  
\label{fig_ha} 
\end{figure*}

In order to investigate the star formation history in NGC\,3603, we have
used  the Wide Field Camera 3 (WFC3) on Hubble Space Telescope (HST) in
the UVIS mode to image NGC\,3603 with  broad band and narrow band
filters as part of  the early release science (ERS) program. In this
paper, we describe the first results of this investigation. The
observations and analysis procedure  are presented in
Section~\ref{sec_obs}. In Section~\ref{sec_cmd} we show the resulting
color magnitude diagram (CMD) and the estimation of field contamination
and extinction. In Section~\ref{sec_ha} we describe  the identification
of pre-Main Sequence (PMS) stars, while their spatial distribution and a
discussion of the preliminary  physical implications of the results can
be found in Section~\ref{sec_ks} and~\ref{sec_conc}, respectively.

\section{Observations and data reduction}
\label{sec_obs}

The photometric data used in this work consist of a series of deep
multi-band images  acquired with the new WFC3 on board of the HST. The
WFC3 consists of two detectors, one optimized for observations in the
wavelength range $\sim200$ to $\sim 1000~nm $ (UVIS channel) and the
other between  $\sim0.9$ and $\sim 1.7~\mu m $ (IR channel).  The  UVIS
detector consists of two 2Kx4K CCDs covering a field of view (FoV) of
162\arcsec x 162\arcsec~at a plate scale of 0\farcs04/px. The IR
detector is a single 1Kx1K HgCdTe CCD offering a total FoV of 123\arcsec
x 136\arcsec~at a pixel resolution of 0\farcs13. A more detailed
description of the WFC3 and its current performances can be found
in~\cite{won10}.

The data used in this work are part of the ERS  observations\footnote{A
complete description of the program, targets and the observations can be
found at the web page http://archive.stsci.edu/prepds/wfc3ers/} obtained
by the WFC3 Scientific Oversight Committee for the study of star
forming  regions in nearby galaxies (Program ID number 11360). NGC\,3603
was observed using both the UVIS and IR channels.  In this paper we use
the images taken through the broad-band $F555W$  and $F814W$ and
narrow-band $F656N$ filters for a total exposure time of  1000s, 1550s
and 990s, respectively. The IR data-set will be presented in a separate
paper (Spezzi et al. 2010, in preparation).

Three images with approximately the same exposure times  were taken with
a few pixel dithering in order to allow for the removal  of cosmic rays,
hot pixels and other detector blemishes.  All the observations were
performed so that the core of NGC\,3603 is roughly at the center of the
camera's FoV.  In Figure~\ref{fig_ha} we show a mosaic of the images in
the $F555W$ (left panel) and  $F656N$ (right panel) filters as obtained
with the PyRAF/MULTIDRIZZLE package.

The photometric analysis of the entire data-set was performed on the
flat-fielded (FLT) images by adopting  the following strategy. The
images, corrected for bias and flatfield, need a further field-dependent
correction factor to  achieve uniformity in the measured counts of an
object across the field. Applying the correction simply involves
multiplying the FLT images by the pixel area map images. A large number
of isolated, well exposed stars were selected in every image over the
entire FoV in order to properly model the point spread function (PSF)
with the DAOPHOTII/PSF routine~\citep {ste87}.  We used a Gaussian
analytic function and a second order look-up table was necessary  in order to
properly account for the spatial variation in the images.

A first list of stars was generated by searching for objects above the
$3\sigma$ detection limit in each  individual image  and a preliminary
PSF fitting run was performed using DAOPHOTII/ALLSTAR. We then used
DAOMATCH  and DAOMASTER to match all stars in each chip, regardless of
the filter, in order to get an accurate coordinate transformation
between the frames. A master star list was created using stars detected
in the $F814W$ band (the deepest in the UVIS data-set)  with the
requirement that a star had to be detected in at least two of the three
images in this filter.

We used the sharpness ($sh$) and chi square parameters given  by
ALLSTAR in order to remove spurious detections.  It has already been
shown~\citep[see e.g.][]{col96,asc07} that these parameters are good
tracers of the photometric quality. By using a sample of real stars we
found the range  $-0.15<sh<0.15$ to be safe enough to eliminate most
spurious objects.  The final catalogue was then obtained by rejecting
any residual spurious source (mostly associated with bright emission
peaks in the HII region not due to point sources) through visual
inspection of the drizzled  images.    The master list was then used as
input for ALLFRAME~\citep{ste94},  which simultaneously determines the
brightness of the stars in all frames while enforcing one set of 
centroids  and one transformation between all images.  All the
magnitudes for each star were normalized to a reference frame and
averaged  together, and the photometric error was derived as the
standard deviation of the repeated  measurements.  The final catalogue 
of the UVIS  $F555W$, $F656N$ and $F814W$ bands contains  around 10,000
stars. 

The photometric calibration was performed following~\cite{ka09}. A
sample of bright isolated stars was used to transform the instrumental
magnitudes to a fixed aperture of  0\farcs4. The magnitudes were then
transformed into the VEGAMAG system by adopting the synthetic zero
points  for the UVIS bands~\citep[see Table 5 of][]{ka09}. Hereafter we
will refer to the calibrated magnitudes as V, I , and H$\alpha$ to 
indicate $m_{F555W}$, $m_{F814W}$ and $m_{F656N}$,  respectively. 

The WFC3/UVIS channel is affected by geometric distortion and a
correction is necessary  in order to properly derive the absolute
positions of individual stars in each catalogue. We used the distortion
coefficients derived by~\cite{koz09} to obtain relative star positions
that are corrected for distortion.   We then used the stars in common
between our UVIS and the 2MASS  catalogues to derive an astrometric 
solution and  obtain the absolute RA and DEC positions of our stars.  We
find a systematic residual of $\sim 0\farcs3$ with respect to the 2MASS
coordinates.

\begin{figure}[t] 
\centering
{\includegraphics[scale=0.44]{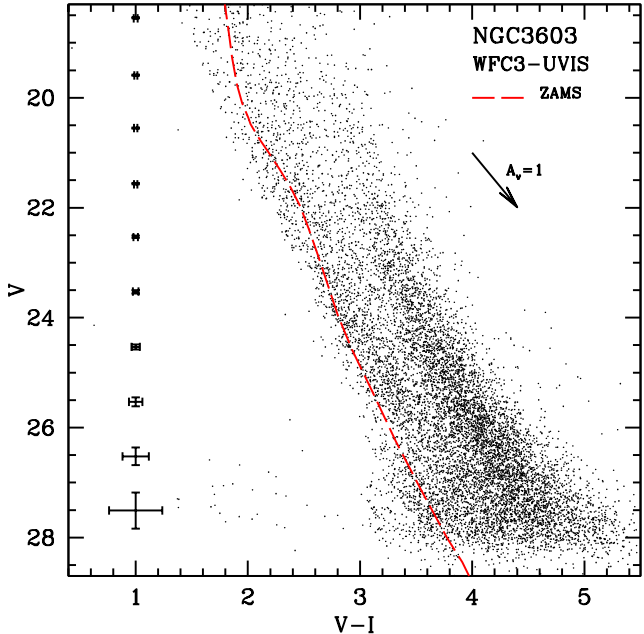}}
\caption{Color magnitude diagram of NGC\,3603 as obtained from WFC3
observations.  The ZAMS from~\citet{mar08} at the solar metallicity
for an assumed distance modulus $(m-M)_0=13.9$ and average extinction
A$_V=4.5$ is over-plotted (dashed line).  The average photometric
magnitudes and color errors are shown (crosses).  The reddening
vector (Av=1) is also shown.} 
\label{fig_cmd} 
\end{figure}

\section{The UVIS Color Magnitude Diagram}
\label{sec_cmd}

The CMD as obtained with our photometric reduction procedure is shown
in  Figure~\ref{fig_cmd}. Typical photometric errors (magnitudes and
colors) are indicated by black crosses. This is the deepest and most
accurate optical diagram derived so far for this cluster, demonstrating
the unprecedented photometric capabilities and the  resolving power of
the new WFC3. 
We show in the CMD the position of the zero age MS (ZAMS)
from~\cite{mar08} for solar metallicity (solid line), having adopted a
distance modulus $(m-M)_0=13.9$ (see HA08 and references therein), the
average value of the extinction A$_V=4.5$ as reported in the
literature~\citep[see][hereafter SB04]{sun04}, and having assumed the
extinction law of \citet{car89}. Note that the value of A$_V=4.5$ is
assumed here only for illustration purposes and for comparison with
previous studies, as we will show later that the actual extinction value
is larger in the area that we studied.  Nonetheless, the position of the 
ZAMS obtained in this way
helps us to identify  a population of candidate MS stars that extend
from the saturation limit at $V \simeq 17$ down  to $V \simeq 26$. As we will show in
Section~\ref{sec_err}  and~\ref{sec_lett}, most of these objects are 
likely foreground field stars but some of them represent a bona-fide low
mass  MS population belonging to the cluster.

A discontinuity in the stellar color distribution in this CMD 
separates a population of objects along the ZAMS from one clearly
grouped at redder colors and consistent with the young population of
PMS stars already detected in NGC\,3603 (see e.g. HA08; SB04). This
confirms that NGC\,3603 is an active star forming region. In order
to learn more about the properties of this recent star formation
episode, it is useful to compare our CMD with PMS isochrones. However,
this requires detailed knowledge of possible sources of error such as
the contamination from field stars and the presence of differential
reddening. We address these issue here below. 

\subsection{Field star contamination and extinction}
\label{sec_err}

As already discussed by HA08  and~\cite{nu02}, the region of  the CMD
fitted by the ZAMS in our CMD is contaminated by field stars with a 
luminosity distribution roughly in agreement with Galactic models.  
In order to quantify the degree of contamination, we
generated a simulated catalogue of field stars using the Galactic model
of \cite{rob03}. The catalogue covers a projected area of
$162\arcsec \times 162\arcsec$, corresponding to the FoV of the WFC3
(see Section~\ref{sec_obs}), with a diffuse extinction of
$0.7$\,mag/kpc in a distance interval of 7\,kpc (i.e. the cluster
distance, see  Section~\ref{intro}). Since the magnitudes of the
synthetic stars  are given in the Johnson-Cousins (JC) photometric
system, we used model atmospheres from the ATLAS9 library of \cite{ku93}
to calculate the magnitude difference between the JC and WFC3
photometric systems as a function of the effective temperature of the
stars. Comparison of the synthetic catalogue with our CMD reveals that
about 85\% of the stars in the region around the ZAMS in our photometry
are potentially field stars, while it decreases to a few percentage towards
the region populated by PMS stars.

Recently,~\cite{roc10} presented a proper motion study of
NGC\,3603  based on HST/WFPC2 (Wide Field Planetary Camera 2)
observations obtained ten years apart, respectively in 1997 and 2007.
On the left panel of their Figure 2 these authors show the position of field
stars (open circles) on the CMD obtained with the Planetary Camera
(PC) data. By comparing this diagram with the same obtained  when
considering  only bona-fide cluster members  (i.e. those with a similar
proper motion; central panel on the same figure), we quantify the field
contamination along the MS in the PC data to be of order 50\%. Although
not explicitly stated in their paper, also \cite{bra99} reach a similar
conclusion, as about 50\,\% of the stars on their original lower MS are
still present after statistical subtraction of a comparison field (see
their Figure\,2). Considering that both the PC dataset used
by~\cite{roc10} and the observations of ~\cite{bra99} sample the core
region of the cluster, where the density of cluster stars is highest, we
conclude that the 85\,\% value of field star contamination that we find
in the external regions from the models  of ~\cite{rob03} is reasonable
and might actually be an overestimate of the true value. We therefore
assume it as an upper limit to the contamination level in this field. 
We will return in Section\,\ref{sec_lett} to the work of ~\cite{roc10}
and ~\cite{bra99} to discuss their implications for the study of the
cluster's stellar population, but we first need to address the issue of
differential extinction in this field. 

An efficient method to quantify
the amount of differential extinction in a star cluster is to use the 
position of a star in the observed CMD to calculate its distance from a
fiducial line (e.g. the ZAMS itself) along the direction of the
reddening vector. This distance would be the resultant of two
components, namely E(V-I) on the abscissa and A$_V$ on the ordinate, and
would give us an estimate of the extinction toward the star
itself~\citep[see an example in][]{pio99}. 

This method works under the assumption that the observed stars are at
the same distance, i.e. belong  to the same system. As discussed above, 
the field star contamination in the range of magnitudes covered by our
data is high. The described method 
can be applied in a reliable way only to bright objects (V$<17$), since these are
very likely cluster members and field star contamination is minimized at these magnitudes.
Although our data cannot be used for this purpose, since all stars
brighter than $V\sim17.5$ are saturated, using shorter exposures SB04
were able to perform a study of differential reddening in NGC\,3603
taking advantage of multi-band HST photometry of the bright massive
stars (i.e. the same objects that are saturated in our images). These
authors were able  to map the variation of E(B-V) as a function of the
distance from the cluster centre (see their Figure 5b). They found that
the value A$_V \simeq 4.5$ is  representative of the very centre of the
OB stars association, while  they noticed an increase toward the
external regions. Following the work of SB04 and adopting $R_V =3.55$ as
they suggest, we estimate that the mean value of A$_V$ in the area
sampled by our observations (from $\sim10\arcsec$ to $\sim70\arcsec$) is
A$_V=5.5$. Therefore, in the rest of this paper we will adopt this value
to correct our magnitudes for extinction. The final CMD corrected in
this way is shown in Figure~\ref{fig_cmd_dered}.

\subsection{The PMS population in NGC\,3603}
\label{sec_lett}

\begin{figure} 
 \centering
  {\includegraphics[scale=0.44]{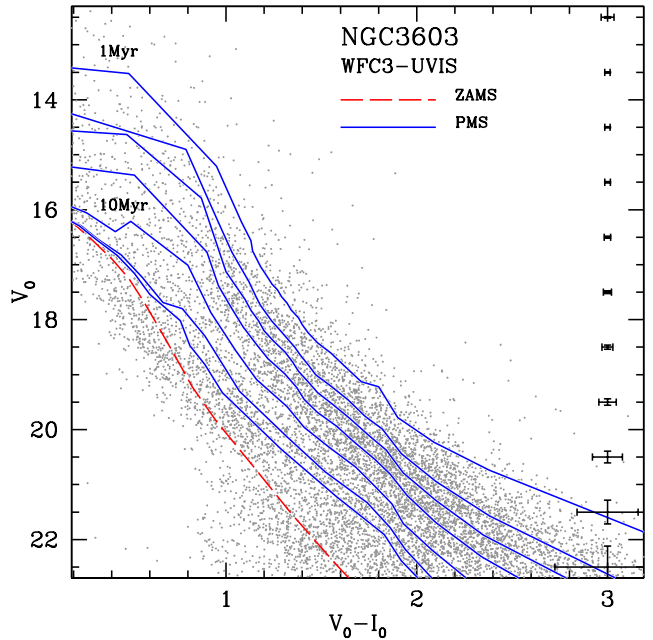}}
   \caption{De-reddened CMD. The ZAMS from 
   ~\cite[][dashed line]{mar08} is used to identify the location of MS stars.
  The population of PMS stars is fitted with 1, 2, 3, 5, 10, 20 and 30 Myr PMS isochrones (from right to the 
left) from~\cite{sie00} (solid lines). The position of the 1 and 10 Myr isochrones is 
indicated. According to the models a mass of $M\sim0.3 M_{\odot}$ is reached
   at magnitude V$=23$.} \label{fig_cmd_dered} 
   \end{figure}

Once extinction is taken into account, the CMD shown in
Figure~\ref{fig_cmd_dered} can be used to determine stellar ages 
through PMS isochrone fitting. PMS isochrones with ages of 1, 2, 3, 10,
20 and 30~Myr from~\cite{sie00} are shown in the figure from right to
left. We used the same value of the distance modulus adopted for the
ZAMS fit. Note that the~\citet{sie00} models are not available for the 
WFC3 photometric system. As above, we used the ATLAS9 library of
\cite{ku93} to calculate the magnitude differences between the  JC to
WFC3 photometric systems. After this correction, the models indicate 
that the lowest mass that we reach for PMS stars is 0.3 M$_{\odot}$.

In the magnitude range $V<20$, where the photometric uncertainty on the
$V-I$ color ($\sim 0.05$) is smaller than the typical isochrone
separation, the CMD suggests for our PMS stars an age in the range from
1 to 10~Myr, with an average age of 3~Myr. Assuming the PMS isochrones are
correct, this can already be considered as tentative evidence of an age spread in
the stellar population of NGC\,3603.

Recently, HA08 studied the IMF of NGC\,3603 using near-infrared (IR)
imaging from ground-based adaptive optics photometry of the cluster
center obtained with the NAOS--CONICA (NACO) camera and the wider field
Infrared Spectrometer and Array Camera (ISAAC) at the Very Large
Telescope (VLT). Their $JHKL'$ bands photometry reaches the magnitude
limit $J\sim20.5$  (i.e. $\sim 0.4$M$_{\odot}$ for cluster stars) in an
area of $\sim 110\arcsec$ radius from the cluster center. By comparing
their CMD to~\cite{ba98} PMS evolutionary models, they identify a
population of PMS stars with ages of $0.5-1.0$\,Myr. Moreover, by adopting
a set of MS isochrones from~\cite{le01}, they provide tentative hints of
the presence of an evolved MS population of $2.0-2.5$\,Myr. Note that this
age estimate is based on the comparison of the MS isochrones with the
position of three massive evolved O stars in their IR CMDs. Coupling
this piece of evidence with the presence of the evolved post-red
supergiant star Sher\,25~\citep{mof83} in the cluster field, HA08
hypothesize a possible age spread in the cluster population suggesting
the presence of two distinct bursts in the star formation history,
separated by $\sim 10$\,Myr. It is interesting to note that
recently~\cite{mel08} have placed Sher\,25 at the same distance as NGC
3603, suggesting a common origin.

The same ISAAC observations analyzed by HA08, were previously used 
by~\cite{sto04} and~\cite{bra99} to study the low-mass stars population
in NGC\,3603. While by inspecting the CMDs shown in their Figure 4 and 3c,
respectively, signatures of the presence of a low MS population 
in the CMDs can not be ruled out, both these papers agree on dating 
a PMS population in the range of $0.5-1$\,Myr.

SB04 published what was by then the deepest optical CMDs based on 
a combination of $UBVI$ and $H{\alpha}$ photometry from the Siding
Spring Observatory (SSO) and archival HST/WFPC2 observations. The SSO
ground-based CMDs sample the brightest massive population. The WFPC2 CMD
samples stars down to $V \simeq 21$. In the very central region of the
cluster ($r<12\arcsec$), the CMD shows the presence of a well defined
population of massive MS stars together with a population of low mass
PMS objects. Apparently only few MS stars are detected in the cluster
center at magnitudes fainter than $V \simeq 19.5$, while in the external
regions ($18\arcsec - 120\arcsec$) stars with masses down to $\sim
1$M$_{\odot}$ are detected both in the MS and PMS regions.  

By using archival WFPC2 observations in the F656N band combined with 
the WFPC2 broad-band photometry, SB04 were able to identify objects with
excess H$\alpha$ emission, which is a signature of the PMS phase. While
the majority of these objects on their CMD occupies a region consistent
with very young PMS isochrones (1~Myr), some stars fall near the ZAMS
(see their Figure 7).  SB04 speculate on a possible spread in age of the
cluster stars, but they warn that it is difficult
to reach a firm conclusion because of the decreasing completeness and
photometric accuracy of their photometry at fainter magnitudes
($V>19$).

A careful examination and comparison of the recent studies on
NGC\,3603 mentioned above shows that they do not exclude the possible
presence of multiple star formation episodes in the recent cluster
history and even of a population of low mass MS stars. However, this
scenario necessarily lacks a clear observational confirmation, since
field contamination remains a crucial point in the study of stellar 
populations in NGC\,3603. An efficient way to overcome these obstacles, 
when observations exist that are separated by a sufficiently large
temporal baseline ($\geq 10 yr$), is the use of proper motions to 
separate cluster stars from foreground and background objects.

As mentioned in Section\,\ref{sec_err}, the proper motions study for the core
region of NGC 3603 by~\cite{roc10} showed signatures 
of at least two star formation epochs in the cluster, 1-2 Myr and 4-5 Myr old, respectively. 
Moreover, the CMD as cleaned-up from the field stars, reveals the presence of 
a low mass MS. The authors conclude that the latter is likely a population of 
objects not belonging to the cluster, that the proper motion technique failed to identify. 
We will offer later in our paper new observational evidence supporting the hypothesis 
that an old stellar population belonging to the cluster is present in the region of the CMD 
occupied by the candidate low MS stars identified by~\cite{roc10}.

As an alternative and independent method to investigate the star formation 
history in NGC\,3603, we have decided to take
full advantage of our new deep $F656N$ band exposures to search for 
objects with excess H$\alpha$ emission, since this feature is a good 
indicator of the PMS stage and therefore of recent star formation. By 
looking at the spatial distribution and age of all the objects with
H$\alpha$ excess emission, we will be able to better understand their
cluster membership.

\section{H$\alpha$ emission stars}
\label{sec_ha}

The presence of a strong H$\alpha$ emission line (EW$_{H\alpha}
\gtrsim$10~\AA) in young stellar objects is normally interpreted as a
signature of the mass accretion process onto the surface of the object
that requires the presence of an inner disk \citep[see][and reference
therein]{fei99,Whi03}.  

The traditional approach to search photometrically for H$\alpha$
emitters is based on the use of the R-band magnitude as a measure of the
level of the photospheric continuum near the H$\alpha$ line, so that
stars with strong H$\alpha$ emission will have a large $R-H\alpha$
color. However, as discussed in \citet{DeM10}, since the R band is over
an order of magnitude wider than the H$\alpha$ filter, the R-H$\alpha$
color does not provide an accurate measurement of the stellar continuum
level inside the H$\alpha$ band. Thus, while helpful to identify PMS
stars, the $R-H\alpha$ color does not provide an absolute measure of the
H$\alpha$ luminosity nor of the H$\alpha$ equivalent width. This
additional information can be derived using measurements in the
neighboring V and I bands, as recently shown by \citet{DeM10}.  This
method allows us to reliably identify PMS objects actively undergoing 
mass accretion regardless of their age.  Briefly, the method combines V
and I broad-band photometry with narrow-band H$\alpha$ imaging to
identify all stars with excess H$\alpha$ emission and to measure their
H$\alpha$ luminosity and mass accretion rate \citep[see][for more
details]{DeM10}.

   \begin{figure}
    \centering 
    {\includegraphics[scale=0.44]{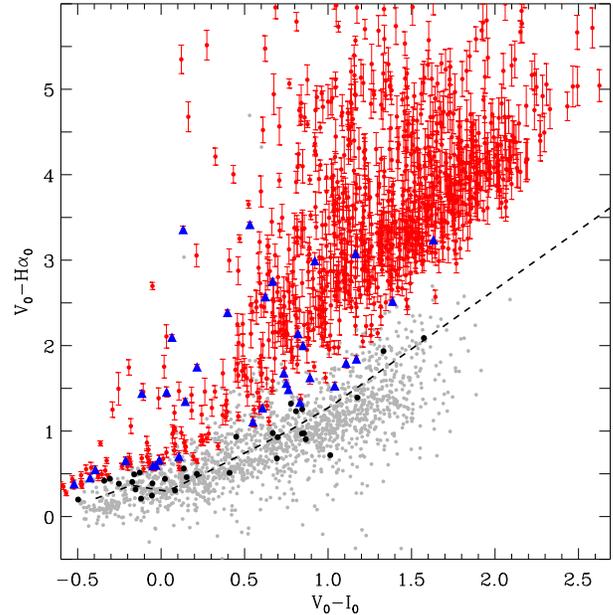}}
   \caption{Selection of H$\alpha$ emission excess stars in a color-color diagram. The dashed line 
represents the median $V-H\alpha$ color representative of stars with no H$\alpha$ excess. 
   Solid triangles and circles show the position
   on the diagram of the counterpart of the 67 objects with H$\alpha$ excess emission by SB04 
   accepted and rejected through our selection criteria, respectively.}  
   \label{fig_hasele}
   \end{figure}

  \begin{figure*}[t] \centering 
   {\includegraphics[scale=0.3]{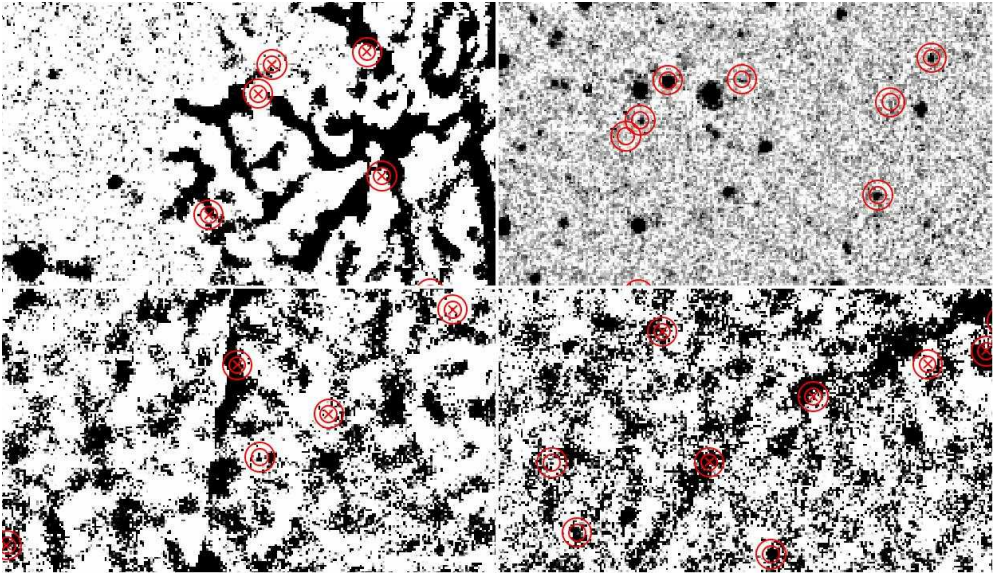}}
   \caption{Four regions (each about 9" x 5" in size) of the drizzled H$\alpha$ image as 
   obtained by applying an unsharp-masking algorithm to highlight and sharpen the 
   details of the dusty and cloudy structures. All candidate H$\alpha$ excess emitters
   are marked with two concentric annuli (4 and 7 pixels radius, respectively). 
 The objects  excluded from our bona-fide sample are marked with a cross.}  
   \label{fig_hasharp}
   \end{figure*}

We followed this approach to select bona-fide PMS stars in the field of
NGC\,3603. Figure~\ref{fig_hasele} shows the $V-H\alpha$ vs. $V-I$  
diagram for the stars in our catalogue. We use the median $V-H\alpha$
dereddened color of stars with small ($<0.05$\,mag) photometric
uncertainties in each of the three V, I and H$\alpha$ bands, as a
function of $V-I$, to define the reference template with respect to
which the excess H$\alpha$ emission is identified (dashed line in
Figure~\ref{fig_hasele}). We selected a first sample of stars with
excess H$\alpha$ emission by considering all those with a $V-H\alpha$
color at least $5\,\sigma$ above that of the reference line, where
$\sigma$ here is the uncertainty on the $V-H\alpha$ color of the star.
Then we calculated the equivalent width of the H$\alpha$ emission line
(EW$_{H\alpha}$) from the measured color excess using Equation~4 of
\citet{DeM10}. We finally considered as bona-fide PMS stars those
objects with EW$_{H\alpha}>$10~\AA~ \citep{Whi03} and $V-I>0$; this
allows us to clean our sample from possible contaminants, such as older
stars with chromospheric activity and Ae/Be stars,
respectively~\citep[see][]{sch07}.

With this approach, we selected a first sample of $\sim 800$
objects with H$\alpha$ excess emission. Through a visual inspection of
the images, we noticed that some of these objects, although well
detected both in the V and I bands, are located along filaments of gas
and dust clearly visible in the H$\alpha$ image (see right panel in
Figure~\ref{fig_ha}).  
It is crucial to consider that, if the centroid of a star falls on top of a 
filament that is only partially included in the annulus that our photometry 
routines use for background subtraction (from 4 to 7 pixel radius), the background will be
underestimated and the derived H$\alpha$ magnitude of the star will be 
over estimated. All sources with H$\alpha$ excess emission have been visually inspected
and all those falling on top of one of such filaments conservatively excluded. 

 As an example, we show in Figure~\ref{fig_hasharp} four regions 
(each about 9" x 5" in size) of the drizzled Halpha image (with the highest spatial 
resolution and free of cosmic rays) as obtained by applying an unsharp-masking 
algorithm to highlight and sharpen the details of the gas filaments. All stars with 
apparent H$\alpha$ excess emission are marked on the Figure with two concentric 
annuli corresponding to the area in which the background has been estimated 
(4 and 7 pixel radius, respectively). We have marked as suspicious and excluded 
from our bona-fide sample all those with significant and non-uniform filament 
contamination inside the photometric aperture (see objects marked with a cross in Figure~\ref{fig_hasharp}).
In this way, we reduced our sample to 412 objects with $5\sigma$ H$\alpha$ 
excess emission that we consider bona-fide PMS stars.

In light of the considerable amount of differential reddening
present in our field, one issue that needs to be taken into account is
the impact that extinction will have on our selection of objects with
H$\alpha$ excess emission. Uncertainties on the extinction will move the
objects in Figure~\ref {fig_hasele}, mostly along the $V-I$ axis,
thereby changing the reference value of the $V-H\alpha$ color. In other
words, variable extinction would change the reference relation (dashed
line) to redder or bluer colors for individual stars. In order to
estimate the impact of differential reddening, we have simulated the
uncertainty introduced by a variation of $\pm 0.2$ mag in $E(V-I)$. 
According to SB04, such a value is the characteristic reddening 
variation seen in the area covered by our observations (see their Figure 5). 
Furthermore, it corresponds to $\Delta Av = \pm 0.5$ mag, which fully covers 
this spread between the $Av =5.5$ value that we find for the reddening and 
the canonical figure for NGC\,3603 ($Av = 4.5$).
If the true $E(V-I)$ value of a star were underestimated, we would also
underestimate its $V-H\alpha$ excess, thereby in practice imposing a
more stringent limit on the excess value itself. By underestimating
$E(V-I)$ by $0.2$\,mag , our $5\,\sigma$ detection limit would correspond
to about $6\,\sigma$.  Conversely, by overestimating the true $E(V-I)$
by $0.2$\,mag, about 17\,\% of the stars detected at the $5\,\sigma$
level will drop below that, but all of them would still be above the
$4\,\sigma$ level. Therefore, while there might be small uncertainties
as to the value of the H$\alpha$ excess for a specific object, the
adoption of an average extinction value across the field  does not
significantly affect our selection of bona-fide PMS stars.

\begin{figure} \centering
  {\includegraphics[scale=0.44]{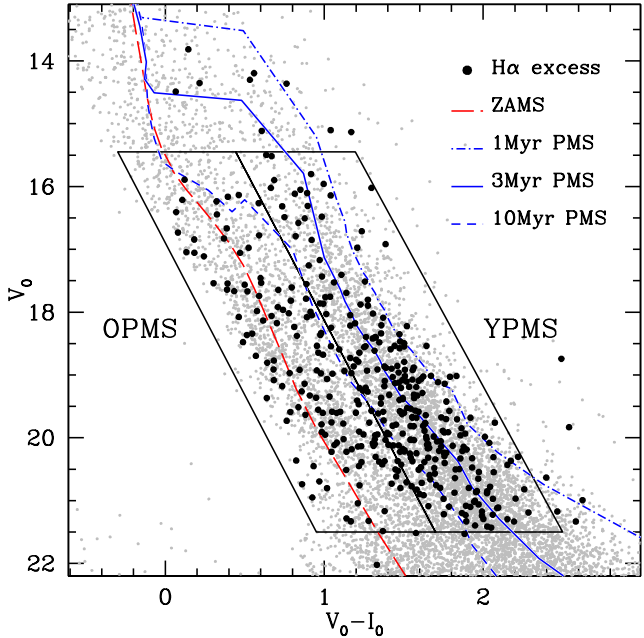}}
   \caption{The location of stars with H$\alpha$ excess (black points) are shown on the CMD 
together with a ZAMS and the 1, 3 and 10~Myr PMS isochrones adopted in Figure~\ref{fig_cmd_dered} 
(long dashed, dotted-dashed, solid and short-dashed line, respectively).
As expected, most of the selected objects are young PMS stars (age $<10$ Myr; YPMS box). 
Surprisingly, a number of sources are located on or near the ZAMS isochrone, i.e. in a position of the CMD 
compatible with a older PMS population (age $> 10$ Myr; OPMS box). The selection boxes of the two populations are shown 
(black lines). }  \label{fig_ha_uv}
   \end{figure}

In Figure~\ref{fig_ha_uv} we indicate with black dots the positions
in the average dereddened CMD of all objects with excess H$\alpha$ emission. The
comparison with the theoretical 1, 3 and 10 Myr PMS isochrones of
\cite{sie00} (dotted-dashed, solid and short-dashed line, respectively) 
suggests a typical age of $\sim 3\,Myr$ for $2/3$ of the
objects with H$\alpha$ excess emission, but a surprising
$1/3$ of them are lying at or near the ZAMS (long dashed line),
thus suggesting a considerably higher age, of order 20--30 Myr. A
10~Myr PMS isochrone seems to efficiently divide the stars with
H$\alpha$ excess emission in two populations, one lying near the ZAMS
and the other in the PMS region.

Even though the fraction of PMS stars showing accretion is known to drop
rapidly with age, \cite{sic10} recently found evidence of stars older
than 10\,Myr still undergoing mass accretion \citep[see also][and
references therein]{ngu09}. In the LMC, \citet{DeM10} found about
130 actively accreting PMS stars with a median age of 14\,Myr
indicating that $\sim10$\,Myr old stars can show H$\alpha$
excess emission due to the accretion process. 

However, as already discussed in Section~\ref{sec_err}, the
region of the CMD fitted by the ZAMS track is strongly contaminated by
field stars. It is then crucial to understand whether the objects with
H$\alpha$ excess emission that we see in NGC\,3603 are indeed cluster
members belonging to a previous generation or are just very young field stars. 
We will show in  Section~\ref{sec_ks} the results of a statistical
analysis of stellar membership providing convincing evidence that these
older PMS stars have a different distribution from that of field
objects.

   \begin{figure*}
   \centering
      \includegraphics[width=0.49\textwidth]{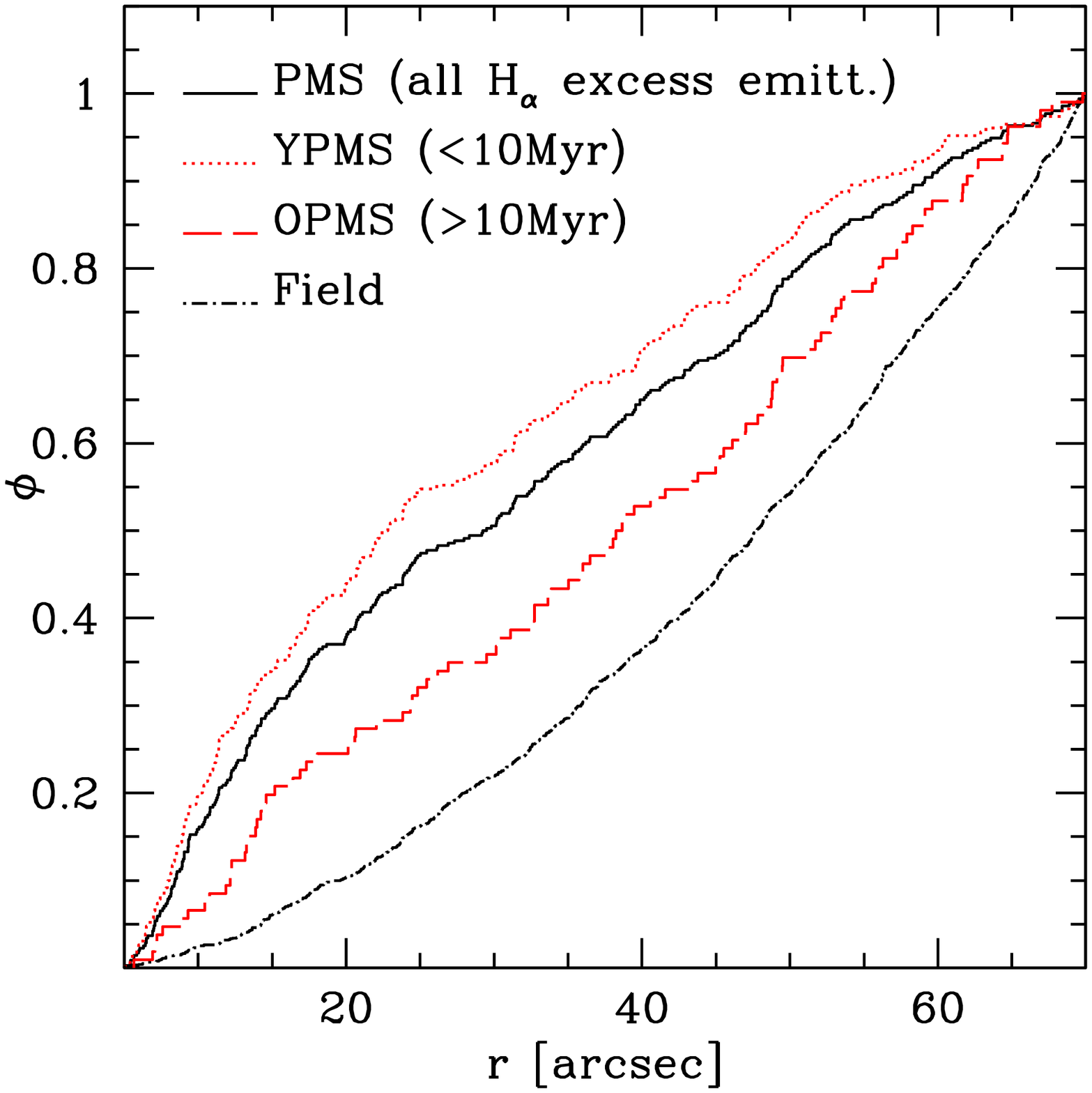}
    \raisebox{1cm}{   \includegraphics[width=0.48\textwidth]{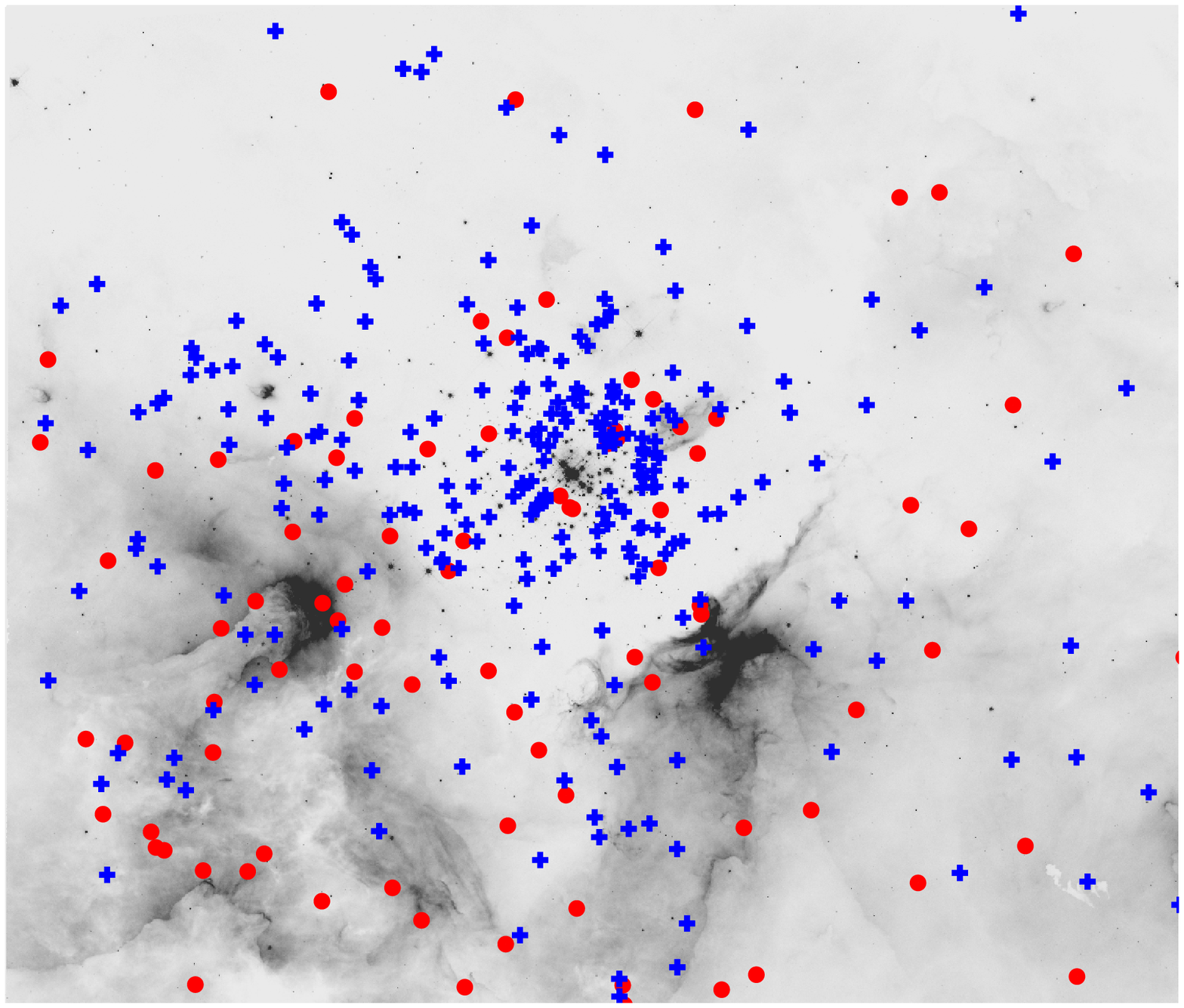}}
   \caption{{\it Left panel}--Cumulative radial distribution of the four star groups as indicated in the legend.
   {\it Right panel}-- The location of YPMS (crosses) and OPMS (solid circles) over-imposed on
   the $F656N$ image already shown on Figure~\ref{fig_ha}.}  \label{fig_ks}
   \end{figure*}

It is known that one property of the accretion process that
characterizes PMS stars is temporal variability, which can be traced to
changes both in the continuum and more importantly in the emission
lines~\citep[e.g.][]{herb86,hart91,ngu09}. As mentioned in
Section~\ref{sec_lett}, SB04 published a list of  sources with H$\alpha$
emission in NGC\,3603 derived from SSO and WFPC2 observations.  We
searched for the counterparts of these objects in our observations by
cross-correlating our photometric catalogue with that of SB04. The stars
detected by SB04 on the SSO data are bright and are all saturated in
our  images. On the other hand, the WFPC2 catalogue of SB04 contains 96
sources selected on the basis of their H$\alpha$ index, of which 67 have
a counterpart in our catalogue. As for the 29 missing objects, ten fall
in the inner $10\arcsec$ of the cluster center, where saturation in our
images makes star detection very difficult. Two of the remaining 19
sources fall outside the UVIS FoV, while the others are very close to
the saturation limit in our catalogue or fall into the gaps of the UVIS
mosaic.    

All these 67 sources had H$\alpha$ excess emission at the time of
the SB04 observations, but only 35 of them have H$\alpha$ excess
emission at the $> 5\,\sigma$ level when our observations were taken.
This number decreases to 23 if we consider only stars with EW$_{H\alpha}
> 10$\,\AA. When lowering the threshold to $3\,\sigma$, the number of
stars with H$\alpha$ excess in common with SB04 grows to 41, but only 25
of them have EW$_{H\alpha}>$10~\AA. This means that most ($\sim
65\,\%$) of the  sources showing H$\alpha$ excess in the study of SB04
(March 1999) do not show it  at the epoch of our observations (Aug
2009). This is in line with the current understanding of the accretion
mechanism whereby the in-falling material from the circumstellar disk is 
subject to bursts, corresponding to peaks in the H$\alpha$
emission~\citep[e.g.][]{fer95}.

SB04 looked for matches between bright X-ray sources and objects
with H$\alpha$ excess emission by using archive Chandra X-Ray
Observatory  images. Only $\sim1/3$ of the 96 H$\alpha$ excess
emission sources  in the WFPC2 data set have a X-ray emission. 
We found that of the 23 stars with H$\alpha$ excess emission in 
common between our catalogue and that of SB04, a total of 
8 have X-ray emission, according to SB04.
While it is expected PMS stars actively undergoing mass accretion   to be
detected in X-rays, it is also true that  the physical processes
related to X-ray flaring and those pertaining to mass accretion are not
the same~\citep[see e.g.][]{fei05}.  As shown by SB04, we should not
expect a complete match between the sources  that are bright in X rays
and those showing H$\alpha$ excess at any given time.  And, obviously,
even less so when the observations are not simultaneous, as in the case
of the Chandra and HST data used by there authors, because of the 
considerable variability to which the accretion process is subject (see
above).

\section{Spatial distribution of PMS stars}
\label{sec_ks}

The presence of stars with H$\alpha$ excess emission overlapping
the ZAMS in the CMD of Figure~\ref{fig_ha_uv} offers new support to the
hypothesis of a spread in the age of the PMS stars in NGC\,3603 (see HA08
and references therein), since this is the region where PMS stars
older than 10\,Myr are expected to be located. One obvious issue to
address is whether these objects are cluster members, as we know that
there is a potentially significant contribution from field stars in this
region (see Section~\ref{sec_err}). In order to investigate cluster
membership for stars showing H$\alpha$ emission, we can look at their
spatial distribution compared to that of field stars.

Based on the distribution of H$\alpha$ excess objects in the CMD,
we define two regions containing bona-fide PMS stars (i.e. those with 
H$\alpha$ excess of different ages). The 10\,Myr PMS isochrone is used as
a guide to define a rough separation between the young PMS population
($<10$\,Myr; hereafter YPMS) and the old PMS population ($>10$\,Myr;
hereafter OPMS). We finally define as field population all stars lying
in the OPMS box and not showing H$\alpha$ excess. The corresponding
selection areas are shown as boxes in Figure~\ref{fig_ha_uv}. It is
important to underline here that, having defined as bona-fide PMS stars
only the objects with excess H$\alpha$ emission, we are in practice
setting a lower limit to the actual number of stars in the PMS phase in
NGC\,3603, since some of them can be in the PMS stage without showing
any H$\alpha$ excess because of the variability of their H$\alpha$ flux
(see Section~\ref{sec_ha}). This is certainly the case for the largest
majority of stars that lie to the right of the 10\,Myr isochrone in
Figure\,3. On the other hand, since we are interested in the radial
distribution of PMS stars, selecting only objects with H$\alpha$ excess
emission does not affect the significance of our  statistical analysis.

The cumulative radial distribution of the four groups defined above
(PMS, YPMS, OPMS and Field) with respect to the cluster center is shown
in the left panel of Figure~\ref{fig_ks}, whereas in the right panel 
the positions of YPMS stars (crosses) and of OPMS stars (filled
circles)  are shown on the $F656N$ band image. To calculate the radial
distribution of these objects we adopted the RA(J2000) = $11^{\rm h}\, 
15^{\rm m}\, 7\fs26$ and DEC(J2000) = $-61\degr\, 15\arcmin\, 37\farcs9$
as coordinates for the cluster center, following SB04. We exclude from
our analysis the innermost $5\arcsec$ radius where there is a high 
concentration  of bright O-B type stars that are saturated  in our long
exposures, and the high crowding level makes object detection difficult.
The radius $r=70\arcsec$ defines the largest circle inscribed in the
FoV.

The graph on the left panel of Figure~\ref{fig_ks} clearly shows that
PMS stars (solid line) are more centrally concentrated than field stars
(dot-dashed line). Furthermore, the radial distributions of YPMS and
OPMS stars (dotted and dashed lines, respectively) are clearly different
from  the field population, supporting the idea that these stars belong
to NGC\,3603. We used a Kolmogorov-Smirnov (KS) test to check the
statistical significance of the differences in the observed
distributions. The test yields more than $3\sigma$ confidence level that
field stars have a different radial distribution from that of the PMS,
YPMS and OPMS groups.  

Interestingly, YPMS stars appear to be more  centrally concentrated than
OPMS objects, contrary to what one would expect in a triggered star
formation scenario. Note that the KS test, as used here, indicates the
probability that the different groups are drawn from the same population
on the basis of their radial distribution with respect to a common
center. In principle, however, the assumption of a common center  of
gravity for the PMS population as a whole could be incorrect, since  the
center of gravity of the OPMS stars might as well differ from that  of
the younger YPMS population. Nonetheless, the analysis presented here 
clearly shows how powerful the use of the H$\alpha$ excess information
can be to identify bona-fide PMS stars and to properly separate them
from the field star population.

\section{Discussion and conclusions: the star formation history of NGC\,3603}
\label{sec_conc}

The literature on the study of the stellar population of NGC\,3603
offers a wide debate on the age of the stars  in this cluster and on
their formation history. While~\citet{sto04} and SB04 suggest 1~Myr  as
a common age for massive MS stars and PMS objects, possible evidence of
larger age spreads comes from the analysis of the population of massive
stars. HA08 derive an age of $2-2.5$\,Myr for MS stars through isochrone
fitting of three massive stars in the  center of the cluster classified
as  WN6h+abs objects by~\citet{cro98}, while they date the PMS stars in
the range $0.5 - 1.0$\,Myr. \citet{hen08} also notice that the
massive-star population in NGC\,3603 appears to be predominantly coeval 
(with an age of $1-2$\,Myr),  but that Sher\,25, and one O-type
supergiant, are likely to be slightly older
\citep[$\sim$4-5~Myr;][]{mel08,  cro06}. Sher\,25  ~\citep[Sher
1965;][]{mof83} is clearly visible in Figure~\ref{fig_ha} and has been
recently  classified as a blue supergiant (BSG) surrounded by an
asymmetric, hourglass-shaped circumstellar nebula by \cite {hen08}.

   \begin{figure*} \centering
      \includegraphics[width=0.49\textwidth]{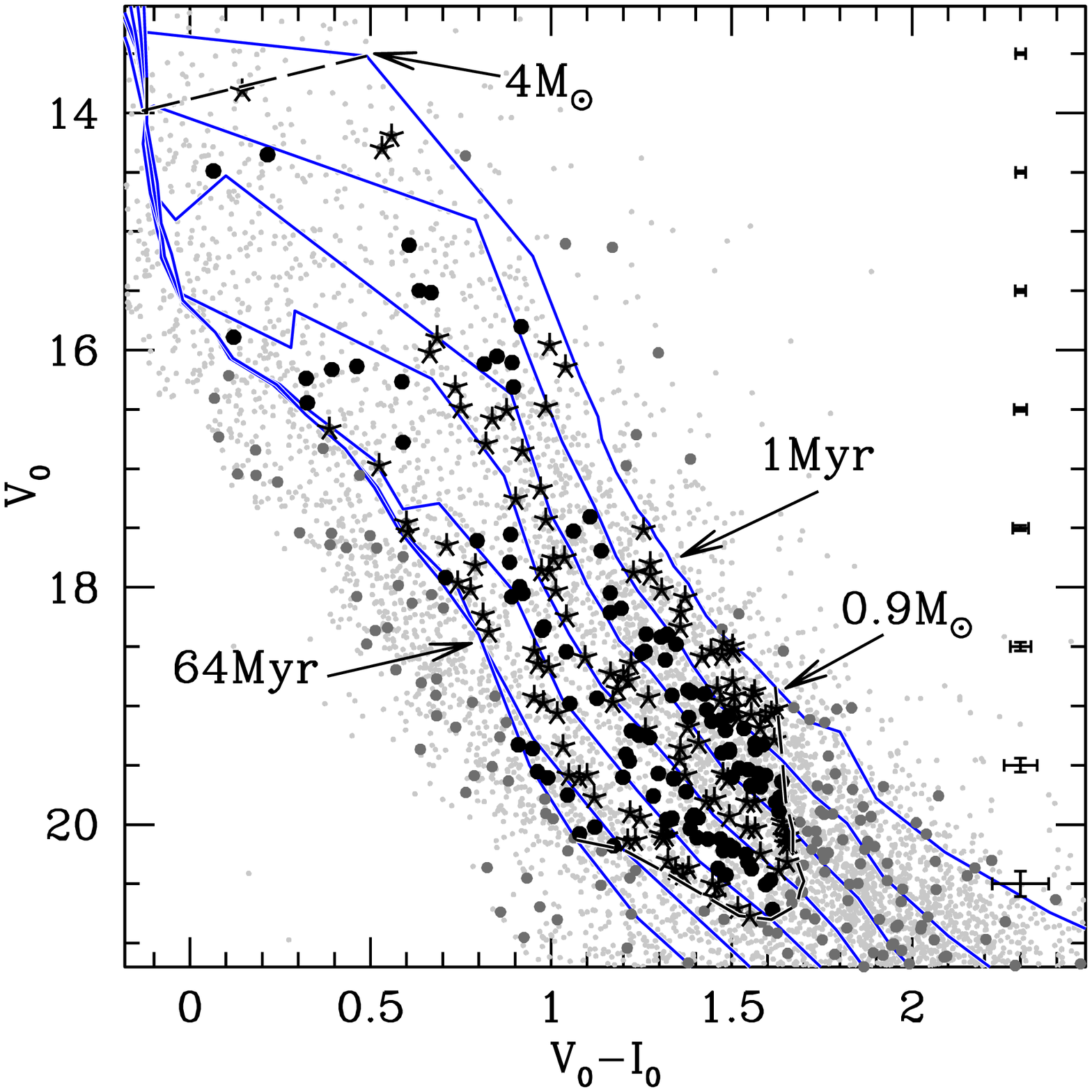}
      \includegraphics[width=0.49\textwidth]{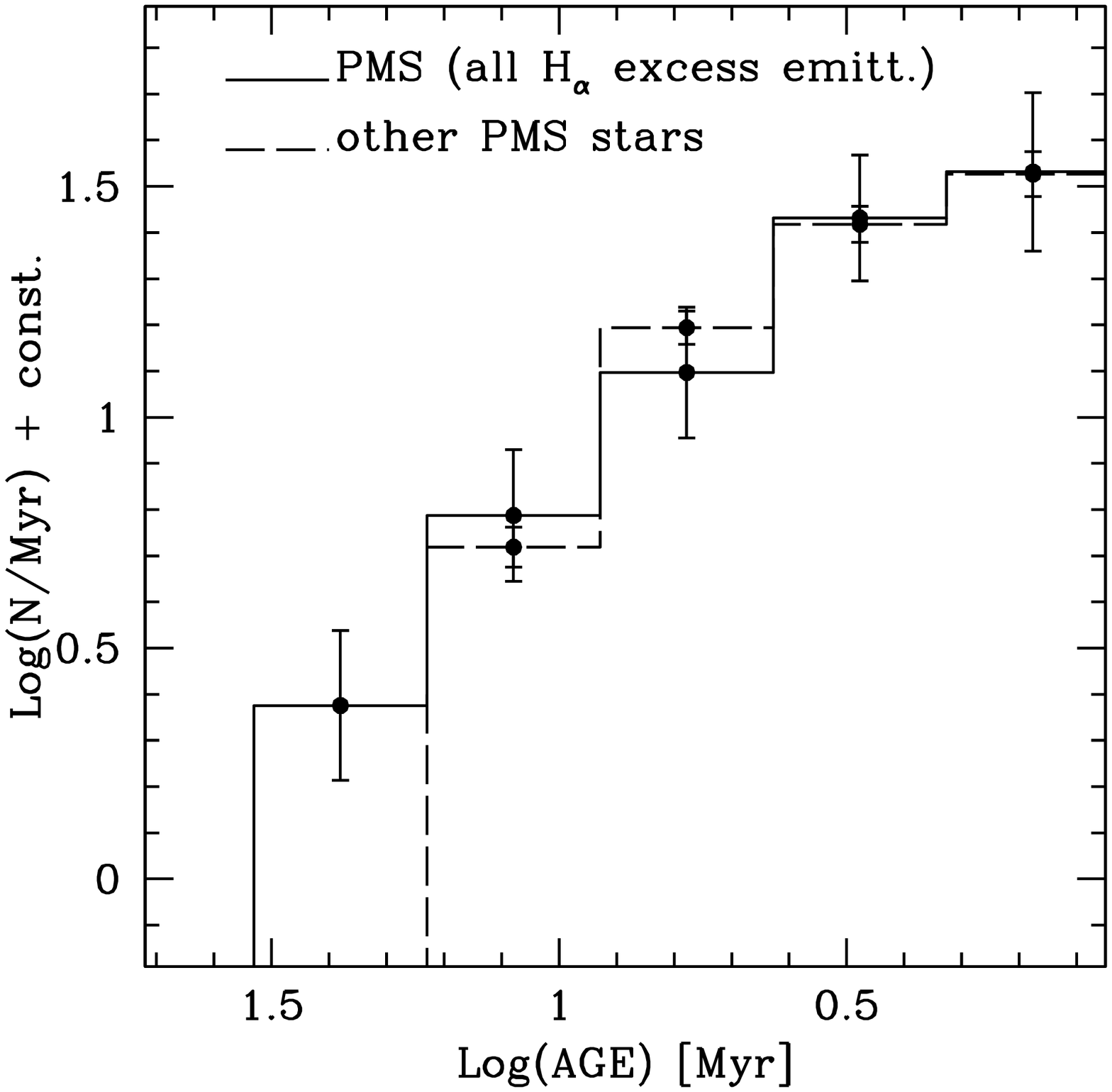}
   \caption{{\it Left panel}-- CMD showing the position of H$\alpha$ stars (solid points and asterisks) together  with the 1, 2, 4, 8, 16, 32 
   and 64 Myr isochrones from~\cite{sie00} used to assign the age to the stars in the range of masses between 0.9 and 4 M$_{\odot}$. 
    PMS stars as counted in different age intervals are shown as thick dark dots
      and asterisks, alternatively. The other PMS stars are shown with large grey dots, whereas all other cluster stars are shown 
      with smaller grey points. 
    {\it Right panel}-- Histogram of the number of PMS stars (i.e. stars showing H$\alpha$ excess emission) per Myr as a function of age (solid line). The same histogram for all other stars is shown for comparison (dashed line).  Note that for the these stars we have only sampled the age distribution up to 16\,Myr, in order to avoid the significant field star contamination in the bluest part of the CMD.}  \label{fig_hist}
   \end{figure*}

HA08 suggests the possibility that, if Sher\,25 belongs to the cluster,
two distinct star formation episodes separated by $\sim 10$\,Myr should
be present in  NGC\,3603. Sher 25 is not the only BSG in the NGC
3603  region. Spectroscopy by~\cite{mof83} revealed two other BSGs in
the vicinity  of the cluster core~\citep[see also Figure 1
from][]{bra97}. The latter authors conclude that the simultaneous
presence of BSGs and stars of MK type O3 V  requires at least two
distinct episodes of star formation  in NGC\,3603 separated by $\sim 10$
Myr.

Our study shows that the OPMS population belongs to NGC\,3603 and not to
the field, thus establishing unambiguously for the first time that
the star formation of this cluster is not characterized by a 1-2 Myr single burst. 
These objects might indeed represent the low mass population of a star
formation episode that occurred more than 10\,Myr ago, which would
naturally explain the presence in the same region of evolved massive
stars (age $> 10$\,Myr) like Sher\,25.

In order to better understand how star formation has proceeded
recently in NGC\,3603, we can look at the age distribution of PMS stars
in the CMD. Shown in the left panel of Figure~\ref{fig_hist} are the PMS
isochrones of~\cite{sie00} for ages of 1, 2, 4, 8, 16, 32 and 64 Myr.
Note that the photometric uncertainty of our data (see large crosses on the figure) 
confirms that it is possible to assign relative ages 
to these stars with an accuracy of better than a factor of 2.  
Solid thick points and asterisks in the figure represent all bona-fide
PMS objects (i.e. those with H$\alpha$ excess emission) having ages in
the range $1 - 64$\,Myr and masses between $0.9$ and 4\,M$_\odot$. The
corresponding age distribution is shown by the solid histogram in the
right panel of the figure. In the same panel we also show, as a dashed
line, the histogram of the age distribution of all stars with no
H$\alpha$ excess younger than 16\,Myr and with masses in the range from
$0.9$ and 4\,M$_\odot$, marked as small dots in the left-hand panel
(note that the  few remaining bona-fide PMS stars outside of these age
and mass ranges, shown as thick grey points, and are not considered
here). For the objects with no H$\alpha$ excess emission we have only
sampled the age distribution younger than 16\,Myr, in  order to avoid the
significant field star contamination in the bluest  part of the CMD.
For comparison purposes, their histogram has been normalized vertically
(i.e. shifted by $1.1$ dex) so as to match the distribution of the
bona-fide PMS stars at the youngest age. 

The histograms of Figure\,\ref{fig_hist} show that star formation in
NGC\,3603 has been ongoing for at least 10--20 Myr  and no gaps are
evident, at least at the level of resolution that we have adopted for
the age (a factor of two, as shown by the size of the bins). 
If we consider stars with no $H\alpha$ excess emission, for which no 
selection effects are present other than photometric completeness that is 
nonetheless always $> 85\,\%$, we would have to conclude that over the 
past $\sim 16$\,Myr the star formation rate in this region has been progressively 
increasing. However, it is important to consider that many of the older stars 
might have migrated out of our FoV. According to~\cite{roc10}, the velocity 
dispersion of stars in the central regions of NGC\,3603 is $\sim 4.5$ km s$^{-1}$ 
and appears to be pretty constant in the mass range that they sample 
($\sim1.7-9M_{\odot}$). This implies that a 10\,Myr old star would have 
had time to travel as far as 45\,pc away from its birthplace, i.e. well beyond 
the $\sim 5 \times 5$\,pc$^2$ area covered by our observations. This might 
be one of the causes of the observed drop in the number of stars with increasing 
age shown in the histogram of Figure\,\ref{fig_hist}, and would also explain 
the somewhat different radial distributions of old and young PMS stars seen 
in Figure~\ref{fig_ks}. A survey of a wider area around NGC\,3603 is needed 
to properly address the evolution of the star formation rate in this cluster.

Interestingly, the age distribution of bona-fide PMS stars
does not seem to differ in any systematic way within the error bars from
that of objects with no H$\alpha$ excess, seemingly suggesting that the
ratio of PMS stars with and without H$\alpha$ excess emission is not a
function of age. This might appear at odds with common wisdom suggesting
that the efficiency of the accretion process at the origin of the
H$\alpha$ emission should decrease with time as a PMS objects approaches
the MS ~\citep[see][and references therein]{sic10,DeM10}. However,
selection effects here can be important. In particular, as PMS objects
approach the MS, both their H$\alpha$ and bolometric luminosities
decrease, but not necessarily in the same way. If the bolometric
luminosity drops more rapidly than the H$\alpha$ luminosity, PMS stars
of older ages will become easier to identify for our method  since it
requires an excess emission at the $5\,\sigma$ level or
higher~\citep[see][]{DeM10}. 

This effect can have important implications on our understanding of the
accretion process and of the star formation rate, which we will address
in a forthcoming paper comparing the star formation process
in a number of young clusters.
We have already started to study the stellar population in the
30\,Doradus  region in the LMC, applying the same observational
strategy  described in this work, and using the recently released WFC3
observations of  this region. By searching for objects with H$\alpha$
excess emission, we have  already found tantalizing evidence of multiple
star formation episodes  (De Marchi et al., in preparation), like in
NGC\,3603. Furthermore, two  distinct  populations with ages of $\sim 1$
and $\sim 15$\,Myr are present in  the population  of the star cluster
NGC\,346 in the Small Magellanic Cloud (De  Marchi et al., in
preparation). Given the different environmental conditions and chemical
compositions of the three clusters ($Z_\odot$ for NGC\,3603, $\sim
1/3~Z_\odot$ for 30 Dor and $\sim 1/10~Z_\odot$ for NGC\,346; see HA08,
Andersen et al. 2009 and Hennekemper et al. 2008,  respectively), this
similarity supports the hypothesis that continuing star formation could
be the preferential channel for the formation of stars in starburst
clusters.

According to~\cite{vin09}, the young cluster Sandage\,96 in NGC 2403 is
known to positively exhibit a young population  ($10-16$\,Myr) together
with a relatively old one ($32-100$\,Myr)  thus suggesting multiple star
formation events in a range of ages at least 4 times wider than in
NGC\,3603. A spread in the MS turn-off has been reported for clusters of
intermediate age in the LMC and has been interpreted as an age spread of
$\sim 300$\,Myr \citep[see][]{mil09}. Moreover, the discovery of
multiple stellar populations along the MS and red giant branch of a
large number of Galactic globular clusters~\citep[][]{pio08,lee09}
requires two or more bursts in the star formation history of these
objects, separated by at a least few $10^7$\,yr~\citep[see][and references
therein]{car10}. 

Therefore, it appears that multiple generations of stars spread over a
wide  range of ages are present in star clusters. A detailed
investigation will be  necessary in order to understand what is at the
origin of the observed age  spread and, for example, what is the
influence of the environment and  chemical/physical state (metallicity,
turbulence, density, mass, etc.) of the  parent molecular cloud on the
formation of stars in clusters.

Establishing whether age spreads like those seen in NGC\,3603 are common
in starburst clusters will have profound implications for theories of
star cluster formation, for the meaning and determination of the IMF and
finally for the general assumption that clusters are simple stellar
populations.

\acknowledgments

We are indebted to an anonymous referee for valuable comments and
suggestions that have helped us to improve the presentation of our work.
We thank Vera Kozhurina-Platais for providing FORTRAN codes for WFC3
geometric distortion corrections and Max Mutchler for producing the
drizzled images shown in Figure~\ref{fig_ha} and~\ref{fig_ks}.
This paper is based on Early Release Science observations made by
the WFC3 Scientific Oversight Committee.  We are grateful to the
Director of the Space Telescope Science Institute for awarding
Director's Discretionary time for this program.  Finally, we are
deeply indebted to the brave astronauts of STS-125 for rejuvenating
HST.

%% in parentheses, after the keyword, but they will not be verified.

{\it Facilities:} \facility{HST (WFC3)}.


%FIGURE%%%%%%%%%%%%%%%%%%%%%%%%


\begin{thebibliography}{}


\bibitem[Alencar \& Batalha(2002)]{al02} Alencar, S.~H.~P., \& Batalha, C.\ 2002, \apj, 571, 378 

\bibitem[Alencar et al.(2005)]{al05} Alencar, S.~H.~P., Basri, G., Hartmann, L., \& Calvet, N.\ 2005, \aap, 
440, 595 

\bibitem[Andersen et al.(2009)]{an09} Andersen, M., Zinnecker, H., Moneti, A., McCaughrean, M.~J., Brandl, 
B., Brandner, W., 
Meylan, G., \& Hunter, D.\ 2009, \apj, 707, 1347 

\bibitem[Appenzeller \& Mundt(1989)]{ap89} Appenzeller, I., \& Mundt, R.\ 1989, \aapr, 1, 291

\bibitem[Ascenso et al.(2007)]{asc07} Ascenso, J., Alves, J., Beletsky, Y., \& Lago, M.~T.~V.~T.\ 2007, 
\aap, 466, 137

\bibitem[Baraffe et al.(1998)]{ba98} Baraffe, Chabrier, Allard, Hauschildt, 1998, \aap, 337, 403

\bibitem[Batalha et al.(2001)]{bat01} Batalha, C., Lopes, D.~F., \& Batalha, N.~M.\ 2001, \apj, 548, 377 

\bibitem[Baumgardt \& Kroupa(2007)]{bau07} Baumgardt, H., \& Kroupa, P.\ 2007, \mnras, 380, 1589

\bibitem[Bertout(1989)]{ber89} Bertout, C.\ 1989, \araa, 27, 351 

\bibitem[Brandl et al.(1999)]{bra99} Brandl, B., Brandner, W., Eisenhauer, F., Moffat, A.~F.~J., Palla, F., 
\& Zinnecker, H.\ 1999, \aap, 352, L69

\bibitem[Brandner et al.(1997)]{bra97} Brandner, W., Grebel, E.~K., Chu, Y.-H., \& Weis, K.\ 1997, \apjl, 
475, L45 

\bibitem[Cardelli et al.(1989)]{car89} Cardelli, J.~A., Clayton, G.~C., \& Mathis, J.~S.\ 1989, \apj, 345, 
245 

\bibitem[Carretta et al.(2010)]{car10} Carretta, E., 
Bragaglia, A., Gratton, R., Recio-Blanco, A., Lucatello, S., D'Orazi, V., 
\& Cassisi, S.\ 2010, arXiv:1003.1723 

\bibitem[Cool et al.(1996)]{col96} Cool, A.~M., Piotto, G., \& King, I.~R.\ 1996, \apj, 468, 655

\bibitem[Crowther et al.(2006)]{cro06} Crowther, P.~A., Lennon, D.~J., Walborn, N.~R., \& Smartt, 2006, 
arXiv:astro-ph/0606717 

\bibitem[Crowther \& Dessart(1998)]{cro98} Crowther, P.~A., \& Dessart, L.\ 1998, \mnras, 296, 622 

\bibitem[De Marchi et al.(2010a)]{DeM10} De Marchi, G., Panagia, N., \& Romaniello, M. 2010, \apj, in press 
(arXiv:1002.4864)

\bibitem[Feigelson \& Montmerle(1999)]{fei99} Feigelson, E.D., \& Montmerle, T. 1999, Ann. Rev. \aa 37, 363

\bibitem[Feigelson(2005)]{fei05} Feigelson, E.~D.\ 2005, 13th 
Cambridge Workshop on Cool Stars, Stellar Systems and the Sun, 560, 175 

\bibitem[Fernandez et al.(1995)]{fer95} Fernandez, M., Ortiz, E., Eiroa, C., \& Miranda, L.~F.\ 1995, \aaps, 
114, 439 

\bibitem[Grebel(2004)]{gre04} Grebel, E.~K.\ 2004, The 
Formation and Evolution of Massive Young Star Clusters, 322, 101 

\bibitem[Harayama et al.(2008)]{har08} Harayama, Y., Eisenhauer, F., \& Martins, F.\ 2008, \apj, 675, 1319 
(HA08)

\bibitem[Hartigan et al.(1991)]{hart91} Hartigan, P., Kenyon, 
S.~J., Hartmann, L., Strom, S.~E., Edwards, S., Welty, A.~D., 
\& Stauffer, J.\ 1991, \apj, 382, 617

\bibitem[Hendry et al.(2008)]{hen08} Hendry, M.~A., Smartt, S.~J., Skillman, E.~D., Evans, C.~J., Trundle, 
C., Lennon, D.~J., Crowther, 
P.~A., \& Hunter, I.\ 2008, \mnras, 388, 1127 

\bibitem[Hennekemper et al.(2008)]{henne08} Hennekemper, E., 
Gouliermis, D.~A., Henning, T., Brandner, W., 
\& Dolphin, A.~E.\ 2008, \apj, 672, 914 

\bibitem[Herbst(1986)]{herb86} Herbst, W.\ 1986, \pasp, 98, 1088 

\bibitem[Johns \& Basri(1995)]{joh95} Johns, C.~M., \& Basri, G.\ 1995, \aj, 109, 2800 

\bibitem[Jones et al.(1981)]{jon81} Jones, T.~J., Ashley, M., Hyland, A.~R., \& Ruelas-Mayorga, A.\ 1981, \mnras, 197, 413

\bibitem[Kalirai et al.(2009)]{ka09} Kalirai, J.~S., et al.\ 2009, Instrument Science Report WFC3 2009-31, 
27 pages, 21 

\bibitem[Kozhurina-Platais et al.(2009)]{koz09} Kozhurina-Platais, V., Cox, C., McLean, B., Petro, L., 
Dressel, L., 
Bushouse, H., Sabbi,E.\ 2009, Instrument Science Report WFC3 2009-33, 22 pages, 33 

\bibitem[Kurucz(1993)]{ku93} Kurucz, R.~L.\ 1993, IAU Colloq.~138:
Peculiar versus Normal Phenomena in A-type and Related Stars, 44, 87 

\bibitem[Larsen(2009)]{lar09} Larsen, S.~S.\ 2009, \aap, 494, 539

\bibitem[Lee et al.(2009)]{lee09} Lee, J.-W., Kang, Y.-W., Lee, J., Lee, Y.-W.\ 2009, Nature, 462, 480 

\bibitem[Lejeune \& Schaerer(2001)]{le01} Lejeune, T., \& Schaerer, D.\ 2001, \aap, 366, 538 

\bibitem[Marigo et al.(2008)]{mar08} Marigo, P., Girardi, L., Bressan, A., Groenewegen, M.~A.~T., Silva, L., 
\& Granato, G.~L.\ 2008, \aap, 482, 883

\bibitem[Mayne \& Naylor(2008)]{may08} Mayne, N.J., \& and Naylor, T. 2008, MNRAS 386, 261

\bibitem[Melena et al.(2008)]{mel08} Melena, N.~W., Massey, P., Morrell, N.~I., \& Zangari, A.~M.\ 2008, 
\aj, 135, 878 

\bibitem[Milone et al.(2009)]{mil09} Milone, A.~P., Bedin, L.~R., Piotto, G., \& Anderson, J.\ 2009, \aap, 
497, 755

\bibitem[Moeckel \& Bate(2010)]{moe10} Moeckel, N., \& Bate, M.~R.\ 2010, arXiv:1001.3417

\bibitem[Moffat(1983)]{mof83} Moffat, A.~F.~J.\ 1983, \aap, 124, 273 

\bibitem[Nguyen et al.(2009)]{ngu09} Nguyen, D.~C., Scholz, 
A., van Kerkwijk, M.~H., Jayawardhana, R., \& Brandeker, A.\ 2009, \apjl, 694, L153 

\bibitem[N{\"u}rnberger \& Petr-Gotzens(2002)]{nu02} N{\"u}rnberger, D.~E.~A., \& Petr-Gotzens, M.~G.\ 2002, 
\aap, 382, 537 

\bibitem[Palla \& Stahler(1999)]{pa99} Palla, F., \& Stahler, S.~W.\ 1999, \apj, 525, 772

\bibitem[Peters et al.(2010)]{pet10} Peters, T., Banerjee, R., Klessen, R.~S., Mac Low, M.-M., Galvan-
Madrid, R., 
\& Keto, E.\ 2010, arXiv:1001.2470

\bibitem[Piotto et al.(1999)]{pio99} Piotto, G., Zoccali, M.,  King, I.~R., Djorgovski, S.~G., Sosin, C., 
Rich, R.~M., 
\& Meylan, G.\ 1999, \aj, 118, 1727

\bibitem[Piotto(2008)]{pio08} Piotto, G. 2008, Memorie della Societa Astronomica Italiana, 79, 334 

\bibitem[Portegies Zwart et al.(2010)]{por10} Portegies Zwart, S., McMillan, S., \& Gieles, M.\ 2010, arXiv:1002.1961 

\bibitem[Robin et al.(2003)]{rob03} Robin, A.~C., Reyl{\'e}, C., Derri{\`e}re, S., \& Picaud, S.\ 2003, 
\aap, 409, 523

\bibitem[Rochau et al.(2010)]{roc10} Rochau, B., Brandner, W., Stolte, A., Gennaro, M., Gouliermis, D., Da Rio, N., Dzyurkevich, N., 
\& Henning, T.\ 2010, \apjl, 716, L90 

\bibitem[Scholz et al.(2007)]{sch07} Scholz, A., Coffey, J., Brandeker, A., \& Jayawardhana, R.\ 2007, \apj, 
662, 1254

\bibitem[Sicilia-Aguilar at al.(2010)]{sic10} Sicilia-Aguilar, A., Henning, T., Hartmann, L.~W.\ 2010, \apj, 
710, 597

\bibitem[Siess et al.(2000)]{sie00} Siess, L., Dufour, E., \& Forestini, M.\ 2000, \aap, 358, 593 

\bibitem[Stetson(1987)]{ste87} Stetson, P.~B.\ 1987, \pasp, 99, 191

\bibitem[Stetson(1994)]{ste94} Stetson, P.~B.\ 1994, \pasp, 106, 250 

\bibitem[Stolte et al.(2004)]{sto04} Stolte, A., Brandner, 
W., Brandl, B., Zinnecker, H., \& Grebel, E.~K.\ 2004, \aj, 128, 765 

\bibitem[Stolte et al.(2006)]{sto06} Stolte, A., Brandner, 
W., Brandl, B., \& Zinnecker, H.\ 2006, \aj, 132, 253

\bibitem[Sung \& Bessell(2004)]{sun04} Sung, H., \& Bessell, M.~S.\ 2004, \aj, 127, 1014 (SB04)

%\bibitem[Vesperini at al.(2008)]{ves08} Vesperini, E., McMillan, S., Portegies Zwart, S.\ 2008, IAU 
Symposium, 246, 181  

\bibitem[Vink{\'o} et al.(2009)]{vin09} Vink{\'o}, J., et al.\ 2009, \apj, 695, 619 

\bibitem[White \& Basri(2003)]{Whi03} White, R.J., \& Basri, G. 2003, \apj, 582, 1109
\bibitem[Wong et al.(2010)]{won10} Wong, M.H., Pavlovsky, C., and Long, K. et al., 2010. ÒWide Field Camera 
3 Instrument Handbook, Version 2.0Ó (Baltimore: STScI)

\end{thebibliography}
\end{document}